\documentclass{optica-article}

\journal{opticajournal} % for journals or Optica Open

\articletype{Research Article}

\begin{document}

\title{Optimizing Dynamic Metasurface Antenna Configurations for Direction-of-Arrival and Polarization Estimation Using an Experimentally Calibrated Multiport-Network Model}

\author{Jean Tapie\authormark{1} and Philipp del Hougne\authormark{1,2,*}}

\address{\authormark{1}Univ Rennes, CNRS, IETR - UMR 6164, F-35000, Rennes, France\\
\authormark{2}Department of Electronics and Nanoengineering, Aalto University, 00076 Espoo, Finland}

\email{\authormark{*}philipp.del-hougne@univ-rennes.fr} %% email address is required; see note below about the corresponding author designation

% use {asbstract*} to suppress the copyright line. Copyright information will be added in production

\begin{abstract*} 
Sensing the direction of arrival and polarization of impinging signals is a key prerequisite for beamforming and interference mitigation in modern wireless communication systems. Dynamic metasurface antennas (DMAs) can multiplex direction- and polarization-dependent field information onto a single detector by sequentially switching between programmable configurations. This makes DMAs attractive for joint direction-of-arrival and polarization (DoA-P) estimation with a single radio-frequency chain. Experimental demonstrations have so far relied on random pre-measured configuration sequences because optimizing the configurations requires an accurate forward model of the fabricated DMA. Here, we use an experimentally calibrated model based on multiport-network theory (MNT) to optimize DMA configuration sequences for DoA-P estimation. Our experimentally calibrated MNT model predicts the dual-polarized far-field response of our 96-element DMA for arbitrary admissible configurations, enabling model-based optimization without additional radiation-pattern measurements. We optimize sequences using effective-rank-based surrogate objectives and compare them with random sequences as a function of the sequence length and the noise level. The optimized sequences yield the largest gains in the intermediate-SNR and intermediate-sequence-length regime, where the inverse problem is neither noise-limited nor already solved by random diversity. We also tackle a dual-source scenario involving a jammer and a desired transmitter. Our results illustrate some of the potential in the context of jamming-resilient communications that is unlocked by experimentally calibrated MNT models for fabricated DMAs.
\end{abstract*}

%%%%%%%%%%%%%%%%%%%%%%%%%%  body  %%%%%%%%%%%%%%%%%%%%%%%%%%
\section{Introduction}

High-resolution wireless sensing conventionally relies on arrays of antennas, each attached to a dedicated radio-frequency (RF) chain. Because RF chains are expensive and power-hungry, computational schemes for wireless sensing using only one or a few RF chains have emerged. These schemes multiplex the wireless signals across a diverse set of measurement modes onto a small number of receivers; subsequently, they recover the sought-after information in post-processing from the multiplexed data. Of particular interest to implement the wave-domain multiplexing are metasurfaces which can offer frequency diversity or, in the case of programmable metasurfaces, configurational diversity. Metasurfaces can thus multiplex information across different frequencies or configurations, respectively.

Computational wireless sensing was initially mostly studied for imaging applications based on frequency multiplexing~\cite{hunt2013metamaterial,fromenteze2015computational,gollub2017large}. Then, the emergence of programmable metasurfaces unlocked single-frequency operation based on multiplexing across sequences of random metasurface configurations~\cite{sleasman2015dynamic,li2016transmission,sleasman2020implementation}. The use of optimized instead of random sequences of DMA configurations was theoretically shown to be beneficial; in particular, if only a subset of the information is salient for the task at hand, then end-to-end optimized task-specific configuration sequences can yield significant benefits~\cite{del2020learned,qian2022noise,saigre2022intelligent}. However, a hurdle to deploy optimized configuration sequences in practice was the lack of an accurate forward model mapping DMA configurations to the corresponding radiation patterns. An accurate forward model is a prerequisite for efficient optimization. Experiments based on random configuration sequences (as well as frequency multiplexing) characterized the utilized radiation patterns by near-field scans, without any possibility of knowing the radiation pattern for a DMA configuration that was not measured.

Later, computational wireless sensing was extended to estimating characteristics such as the direction of arrival (DoA) of far-field sources. Initial efforts were based on frequency-diverse metasurfaces~\cite{yurduseven2019frequency,Imani_Conformal_2023,Alamzadeh_exp_conformal_2025}, while more recent efforts leverage configurational diversity offered by a sequence of random configurations of a DMA~\cite{hoang2021single,Alamzadeh_HybridRIS_IntensityOnly_Ghost_2023,alamzadeh2024computational,zhao2025microwave}. However, the same difficulties due to the lack of an accurate forward model remained. Again, experiments based on random configuration sequences (as well as frequency multiplexing) characterized the utilized radiation patterns by near-field scans~\cite{Alamzadeh_exp_conformal_2025,zhao2025microwave}, without any possibility of knowing the radiation pattern for a DMA configuration that was not measured.

\begin{figure*}
    \centering
    \includegraphics[width=\textwidth]{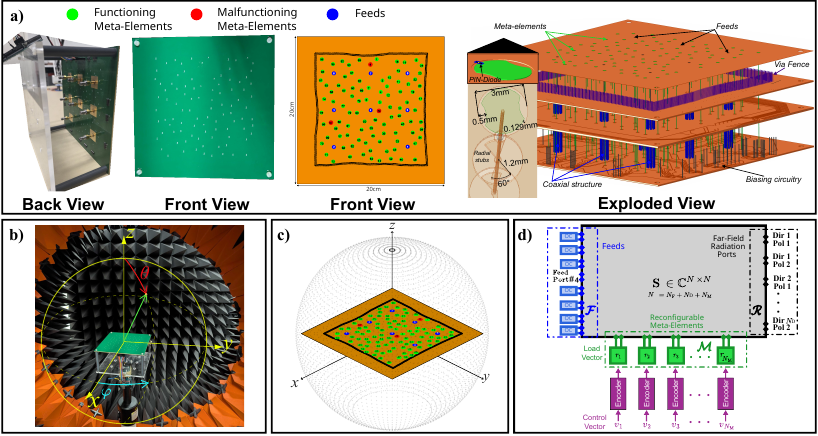}
    \caption{
    Overview of the DMA prototype, measurement setup, and MNT-based system model. 
    (a) Photographs and schematic representation of the chaotic-cavity-backed DMA prototype described in Sec.~\ref{subsec:DMAdesign}. The prototype comprises eight feed ports and 96 1-bit-programmable meta-elements. 
    (b) DMA prototype mounted in the spherical dual-polarization far-field measurement setup. Only the central feed is connected, while the remaining seven feeds are left open-circuited.  
    (c) Discretized far-field grid used for DoA and polarization estimation. Each grid point corresponds to one considered direction; there are two transverse field components ($E_\phi$ and $E_\theta$) for each direction. 
    (d) MNT abstraction of the setup described in Sec.~\ref{sec:system_model}, featuring the DMA's feed ports, the DMA's ``virtual'' ports terminated by tunable loads that represent the tunable lumped elements, and the radiation ports associated with far-field directions and two transverse polarizations.
    }
    \label{Fig1}
\end{figure*}

Recently, we developed a series of techniques for experimentally calibrating an accurate forward model based on multiport network theory (MNT) for any given DMA prototype~\cite{tapie2026experimental,tapie2026channel}. The experimentally calibrated MNT model allows us to accurately predict the radiation pattern associated with any admissible DMA configuration; for a DMA with 96 1-bit-programmable elements such as our prototype shown in Fig.~\ref{Fig1}a, there are $2^{96}$ possible DMA configurations. While we consider the specific DMA prototype shown in Fig.~\ref{Fig1}a, our technique is general thanks to the broad applicability of the MNT approach.

Capitalizing on the experimentally calibrated MNT model for our DMA prototype, we study in this paper the optimization of a sequence of DMA configurations for simultaneous DoA and polarization (DoA-P) estimation in 3D space. We view the computational DoA-P estimation as a first step toward integrated communications, sensing, and wave-domain computing with DMAs~\cite{saigre2023self,10812728}. Indeed, awareness of DoA-P is required not only for beamforming in modern wireless communication systems, but also to identify and null strong interference signals originating from jammers~\cite{yven2025end,hao2025dynamic}.
Beyond enabling the efficient optimization of dedicated DMA configuration sequences, the calibrated MNT model also allows DoA-P estimation to be performed opportunistically when the DMA is configured for another primary purpose, such as communications.

Our contributions are summarized as follows. 
\textit{First}, we formalize problem statements and present corresponding algorithms for single-source and dual-source DoA-P estimation with a single-feed DMA using an experimentally calibrated MNT model as the forward map from configurations to dual-polarized far-field responses. 
\textit{Second}, we use the calibrated MNT model to optimize DMA configuration sequences according to effective-rank-based surrogate objectives, without requiring additional radiation-pattern measurements and without any design-space limitations beyond the constraints of the DMA's physical implementation. 
\textit{Third}, we quantify the resulting DoA-P estimation gains relative to random configuration sequences as a function of the sequence length and the signal-to-noise ratio.
\textit{Fourth}, we demonstrate the feasibility of DoA-P estimation with our DMA prototype in a dual-source scenario with a strong jammer and a weaker desired transmitter.

The remainder of this paper is organized as follows. In Sec.~\ref{sec:system_model}, we describe the MNT system model. In Sec.~\ref{sec:ProblemStatement}, we formulate the single-source and dual-source DoA-P estimation problems. In Sec.~\ref{sec:algorithms}, we present the corresponding DoA-P estimation algorithms. In Sec.~\ref{sec:OptimAlgorithm}, we describe the optimization of the DMA configuration sequence. In Sec.~\ref{sec:Results}, we report our results. We briefly conclude in Sec.~\ref{sec:Conclusion}.

\section{System Model}
\label{sec:system_model}

In this section, we describe the MNT-based mapping from a DMA configuration to the associated far-field radiation pattern. As illustrated in Fig.~\ref{Fig1}d, our MNT-based system model describes the DMA as a single-port ($N_\mathrm{F}=1$) radiating structure parametrized by $N_\mathrm{M}=96$ 1-bit-programmable lumped elements. The DMA's seven unused feeds are open-circuited and treated as part of the static scattering structure; their effect is therefore absorbed into the calibrated effective single-feed MNT model~\cite{tapie2026channel}. 
Each tunable lumped element is treated as a ``virtual'' port terminated by a tunable load, which is justified by the electrically very small nature of the tunable lumped elements~\cite{tapie2026experimental}. The far-field radiation pattern is discretized on a spherical grid of $N_\mathrm{D}$ directions, where each direction is parametrized by the azimuthal angle $\phi \in (-\pi,\pi]$, measured in the $xy$-plane from the positive $x$-axis toward the positive $y$-axis, and the polar angle $\theta \in [0,\pi]$, measured from the positive $z$-axis. For each discretized direction $(\phi,\theta)$, the far field is represented by its two transverse components $E_\phi(\phi,\theta)$ and $E_\theta(\phi,\theta)$ in the spherical basis associated with that observation direction. We thus represent the far field with $2N_\mathrm{D}$ far-field radiation ports.

Altogether, our system model involves $N=N_\mathrm{F}+N_\mathrm{M}+2N_\mathrm{D}$ ports, of which the $N_\mathrm{M}$ ``virtual'' ports are terminated by tunable loads. The static system components constitute an $N$-port network characterized by its scattering matrix $\mathbf{S}\in\mathbb{C}^{N \times N}$. The tunable system components are the ensemble of $N_\mathrm{M}$ tunable individual loads, which constitutes an $N_\mathrm{M}$-port network characterized by its scattering matrix $\mathbf{\Phi}(\mathbf{r})=\mathrm{diag}(\mathbf{r})\in\mathbb{C}^{N_\mathrm{M}\times N_\mathrm{M}}$, where $\mathbf{r} = [r_1,r_2,\dots,r_{N_\mathrm{M}}]^\top\in\mathbb{C}^{N_\mathrm{M}}$ and $r_i\in\mathbb{C}$ is the reflection coefficient of the $i$th tunable load. Throughout this paper, we use a reference impedance of $Z_0=50\ \Omega$ to define all scattering parameters.

The channel vector $\mathbf{h}(\mathbf{r})\in\mathbb{C}^{2N_\mathrm{D}}$ from the single feed port to the $2N_\mathrm{D}$ far-field radiation ports is given by standard MNT~\cite{williams2022electromagnetic,almunif2025network,tapie2026experimental,tapie2026channel} as
\begin{equation}
\mathbf{h}(\mathbf{r})
=
\mathbf{h}_0
+
\mathbf{A}\,
\bigl(
\mathbf{I}_{N_\mathrm{M}}
-
\mathbf{\Phi}(\mathbf{r})\,\mathbf{\Gamma}
\bigr)^{-1}\,
\mathbf{\Phi}(\mathbf{r})\,
\mathbf{b},
\label{eq:MNT}
\end{equation}
where $\mathbf{h}_0=\mathbf{S}_{\mathcal{RF}}\in\mathbb{C}^{2N_\mathrm{D}}$ is the direct channel from the feed port to the far-field radiation ports, $\mathbf{A}=\mathbf{S}_{\mathcal{RS}}\in\mathbb{C}^{2N_\mathrm{D}\times N_\mathrm{M}}$ describes the channels from the ``virtual'' ports to the far-field radiation ports, $\mathbf{\Gamma}=\mathbf{S}_{\mathcal{SS}}\in\mathbb{C}^{N_\mathrm{M}\times N_\mathrm{M}}$ describes the mutual coupling between the ``virtual'' ports, and $\mathbf{b}=\mathbf{S}_{\mathcal{SF}}\in\mathbb{C}^{N_\mathrm{M}}$ describes the channel from the feed port to the ``virtual'' ports. Here, $\mathcal{F}$, $\mathcal{S}$, and $\mathcal{R}$ denote the sets of port indices associated with the feed port, the ``virtual'' ports, and the far-field radiation ports, respectively.

The load vector $\mathbf{r}$ is determined by a control vector $\mathbf{v}\in\{0,1\}^{N_\mathrm{M}}$ based on the two possible reflection coefficients $\alpha\in\mathbb{C}$ and $\beta\in\mathbb{C}$ of the tunable loads~\cite{tapie2026experimental,tapie2026channel}:
\begin{equation}
    \mathbf r (\mathbf{v})
    =
    \alpha \, \mathbf 1
    +
    (\beta-\alpha)\,\mathbf v   \in \mathbb C^{N_\mathrm{M}},
    \label{eq:load_vector}
\end{equation}
where we assume that the two possible reflection coefficients are the same for all loads. Readers interested in generalizations of \eqref{eq:load_vector} to multi-bit tunable elements may refer to~\cite{del2025experimental}.

The system model parameters are thus $\{\alpha,\beta,\mathbf{h}_0,\mathbf{A},\mathbf{\Gamma},\mathbf{b}\}$.
If the geometry and material composition of the DMA prototype were perfectly known and sufficient computational resources were available, then $\{\mathbf{h}_0,\mathbf{A},\mathbf{\Gamma},\mathbf{b}\}$ could be extracted by means of full-wave simulation~\cite{tapie2024systematic,almunif2025network,ramirez2025metasurface}. In practice, however, this route is not feasible for two reasons. \textit{First}, fabrication inaccuracies typically lead to significant deviations from the assumed design, especially for DMA prototypes with strong inter-element coupling~\cite{xu2022extreme,xu2022wide}; yet, strong inter-element coupling is desirable because it maximizes the DMA's wave-domain flexibility~\cite{prod2025mutual,prod2025benefits}. \textit{Second}, the DMA design may be proprietary and hence unknown, or the characteristics of the utilized components may be insufficiently documented. For these reasons, an accurate MNT model for a given DMA prototype must be calibrated experimentally. Yet, since the tunable loads are not connectorized and very numerous, the MNT model parameters can only be estimated indirectly. As discussed in~\cite{tapie2026experimental}, such an indirect experimental estimation of the MNT parameters leads to inevitable ambiguities in the parameters; however, for a consistent set of calibrated proxy parameters, these ambiguities do not affect the mapping from $\mathbf{v}$ to $\mathbf{h}(\mathbf{r}(\mathbf{v}))$ and are hence immaterial for the present work. For notational ease, in this paper we use the same symbols for the true parameters and for the estimated proxy parameters.

\section{Problem Statement}
\label{sec:ProblemStatement}

\subsection{Single Far-Field Source}
\label{subsec:SingleSourceProblemStatement}

We first consider a single source in the DMA's far field. The impinging field can be modeled as a plane wave at the DMA. For simplicity, we assume that the source direction coincides with one of the $N_\mathrm{D}$ discretized directions of the far-field grid introduced in Sec.~\ref{sec:system_model}. We denote the corresponding direction index by $d_0\in\{1,\ldots,N_\mathrm{D}\}$, and the associated angular coordinates by $(\phi_0,\theta_0)$. The source polarization is described in the spherical basis associated with that direction by a vector $\mathbf{c}_0=[c_{\phi,0},c_{\theta,0}]^\top\in\mathbb{C}^{2}$. The norm and global phase of $\mathbf{c}_0$ depend on the source's amplitude and global phase, as well as the propagation loss, and the phase associated with the propagation delay; the influence of these unknown factors is removed in the source's normalized polarization state $\bar{\mathbf{c}}_0=\mathbf{c}_0/\|\mathbf{c}_0\|_2$.

We measure the signal received at the DMA's feed for a sequence of $K$ DMA configurations. The source is assumed to be stationary over the course of these $K$ measurements, meaning that its direction, polarization, amplitude, and phase do not change while the DMA is sequentially configured with the control vectors $\{\mathbf{v}_k\}_{k=1}^{K}$. For the $k$th DMA configuration, the corresponding load vector is $\mathbf{r}_k=\mathbf{r}(\mathbf{v}_k)$ and the far-field channel vector predicted by \eqref{eq:MNT} is denoted by $\mathbf{h}_k=\mathbf{h}(\mathbf{r}_k)\in\mathbb{C}^{2N_\mathrm{D}}$. 
Although \eqref{eq:MNT} was written as a transmit model from the DMA feed to the far-field radiation ports in Sec.~\ref{sec:system_model}, we can also use it for reception by reciprocity. Thus, for the $k$th DMA configuration, the complex signal measured at the feed is
\begin{equation}
y_k
=
h_{\phi,k}(d_0)\,c_{\phi,0}
+
h_{\theta,k}(d_0)\,c_{\theta,0}
+
n_k,
\label{eq:single_source_measurement_scalar}
\end{equation}
where $h_{\phi,k}(d_0)$ and $h_{\theta,k}(d_0)$ are, respectively, the $E_\phi$ and $E_\theta$ components of the entries of $\mathbf{h}_k$ associated with direction $d_0$;  $n_k\in\mathbb{C}$ denotes measurement noise. Collecting the $K$ scalar measurements into a vector, we define $\mathbf{y}
=
[y_1,y_2,\ldots,y_K]^\top
\in\mathbb{C}^{K}$ for notational ease.

The single-source DoA-P estimation problem is summarized as follows: Given the applied DMA configurations $\{\mathbf{v}_k\}_{k=1}^{K}$, the corresponding measured feed signals $\{y_k\}_{k=1}^{K}$, and the calibrated MNT model that maps each $\mathbf{v}_k$ to $\mathbf{h}(\mathbf{r}(\mathbf{v}_k))$, estimate the source direction index $d_0$ and the normalized source polarization vector $\bar{\mathbf{c}}_0$. 

\subsection{Two Far-Field Sources}
\label{subsec:TwoSourceProblemStatement}

We also consider a two-source scenario, motivated by situations in which signals from a strong jammer and a weaker desired transmitter impinge on the DMA from different far-field directions~\cite{yven2025end,hao2025dynamic}. The two sources are assumed to be stationary over the course of the $K$ measurements and to coincide with two discretized directions indexed by $d_1,d_2\in\{1,\ldots,N_\mathrm{D}\}$, with $d_1\neq d_2$. Their polarization-and-amplitude vectors are denoted by $\mathbf{c}_1,\mathbf{c}_2\in\mathbb{C}^{2}$.

For the $k$th DMA configuration, the received feed signal is the superposition
\begin{equation}
y_k
=
h_{\phi,k}(d_1)c_{\phi,1}
+
h_{\theta,k}(d_1)c_{\theta,1}
+
h_{\phi,k}(d_2)c_{\phi,2}
+
h_{\theta,k}(d_2)c_{\theta,2}
+
n_k .
\label{eq:two_source_measurement_scalar}
\end{equation}

The dual-source DoA-P estimation problem is summarized as follows: Given the applied DMA configurations $\{\mathbf{v}_k\}_{k=1}^{K}$, the corresponding measured feed signals $\{y_k\}_{k=1}^{K}$, the calibrated MNT model that maps each $\mathbf{v}_k$ to $\mathbf{h}(\mathbf{r}(\mathbf{v}_k))$, and prior knowledge that the jammer's signal is much stronger than the desired transmitter's signal, estimate the source direction indices $d_1$ and $d_2$ as well as the normalized source polarization vectors $\bar{\mathbf{c}}_1$ and $\bar{\mathbf{c}}_2$.

\section{DoA-P Estimation Algorithms}
\label{sec:algorithms}

In this section, we describe our algorithms for DoA-P estimation in the two estimation problems from Sec.~\ref{sec:ProblemStatement}.

\subsection{Single-Source DoA-P Estimation Algorithm}
\label{subsec:single_source_algorithm}

We estimate the source direction and polarization by a grid search over all candidate directions. For a candidate direction $d\in\{1,\ldots,N_\mathrm{D}\}$, we collect the two response vectors associated with the two transverse polarizations into
\[
\mathbf{D}_d
=
\begin{bmatrix}
h_{\phi,1}(d) & h_{\phi,2}(d) & \cdots & h_{\phi,K}(d)\\
h_{\theta,1}(d) & h_{\theta,2}(d) & \cdots & h_{\theta,K}(d)
\end{bmatrix}^\top
\in\mathbb{C}^{K\times 2}.
\]
For this candidate direction, the best-fitting polarization-and-amplitude vector $\hat{\mathbf{c}}(d)\in\mathbb{C}^{2}$ is obtained by least squares:
\begin{equation}
\hat{\mathbf{c}}(d)
=
\underset{\mathbf{c}}{\mathrm{argmin}}\,
\|\mathbf{y}-\mathbf{D}_d\,\mathbf{c}\|_2^2
=
\mathbf{D}_d^{+}\,\mathbf{y},
\label{eq:single_source_lsq_pol}
\end{equation}
where $(\cdot)^+$ denotes the Moore--Penrose pseudoinverse. The corresponding normalized projection score is
\begin{equation}
\eta(d)
=
\frac{\|\mathbf{D}_d\, \hat{\mathbf{c}}(d)\|_2^2}{\|\mathbf{y}\|_2^2}.
\label{eq:single_source_score}
\end{equation}
The DoA estimate is obtained as
\begin{equation}
\hat d
=
\underset{d}{\mathrm{argmax}}\,
\eta(d),
\label{eq:single_source_doa_estimate}
\end{equation}
and the associated normalized polarization state estimate is
\begin{equation}
\hat{\bar{\mathbf{c}}}
=
\frac{\hat{\mathbf{c}}(\hat d)}
{\|\hat{\mathbf{c}}(\hat d)\|_2}.
\label{eq:single_source_pol_estimate}
\end{equation} 

To summarize, our single-source algorithm consists of a least-squares polarization fit for each candidate direction, followed by a dictionary search over the discretized far-field grid.

\subsection{Dual-Source DoA-P Estimation Algorithm}
\label{subsec:dual_source_algorithm}

For the dual-source scenario, we use the prior knowledge that one source (the jammer) is much stronger than the other (the desired transmitter). This motivates a sequential estimation-and-cancellation strategy. Our algorithm applies the single-source estimator from Sec.~\ref{subsec:single_source_algorithm} twice: \textit{first}, the strong jammer is estimated while treating the weak desired signal as part of the effective noise; \textit{second}, the estimated jammer contribution is subtracted and the desired transmitter is estimated from the residual. In a final step, we refine the polarization estimates via a joint least-squares solve.

We thus begin by applying the single-source algorithm from Sec.~\ref{subsec:single_source_algorithm} directly to the measured vector $\mathbf{y}$, yielding the strongest-source estimate
\[
\hat d_1
=
\underset{d}{\mathrm{argmax}}\,
\eta(d),
\qquad
\hat{\mathbf{c}}_1
=
\mathbf{D}_{\hat d_1}^{+}\,\mathbf{y}.
\]
Then, we subtract the contribution of the strongest source from the measurement vector:
\begin{equation}
\mathbf{y}_\mathrm{res}
=
\mathbf{y}
-
\mathbf{D}_{\hat d_1}\,\hat{\mathbf{c}}_1 .
\label{eq:dual_source_residual}
\end{equation}
Subsequently, we characterize the weaker source by applying the same projection-score search to the residual vector $\mathbf{y}_\mathrm{res}$. Specifically, for each candidate direction $d$, we compute
\[
\hat{\mathbf{c}}_\mathrm{res}(d)
=
\mathbf{D}_d^{+}\,\mathbf{y}_\mathrm{res},
\]
and the corresponding residual projection score
\begin{equation}
\eta_\mathrm{res}(d)
=
\frac{\|\mathbf{D}_d\,\hat{\mathbf{c}}_\mathrm{res}(d)\|_2^2}
{\|\mathbf{y}_\mathrm{res}\|_2^2}.
\label{eq:dual_source_residual_score}
\end{equation}
The second-source direction estimate is then
\begin{equation}
\hat d_2
=
\underset{d}{\mathrm{argmax}}\,
\eta_\mathrm{res}(d),
\label{eq:dual_source_second_doa}
\end{equation}
where we exclude a small angular neighborhood around $\hat d_1$ from the search set to avoid re-detecting the strongest source.

After the two DoAs have been estimated, we perform a joint least-squares refit of the source polarizations with both DoAs fixed. Defining $\mathbf{D}_{12}
=
\begin{bmatrix}
\mathbf{D}_{\hat d_1} & \mathbf{D}_{\hat d_2}
\end{bmatrix}
\in\mathbb{C}^{K\times 4}$, 
we estimate the two polarization-and-amplitude vectors jointly as
\begin{equation}
\begin{bmatrix}
\hat{\mathbf{c}}_1^\top & \hat{\mathbf{c}}_2^\top
\end{bmatrix}^\top
=
\mathbf{D}_{12}^{+}\,\mathbf{y}.
\label{eq:dual_source_joint_refit}
\end{equation}
The normalized polarization state estimates are finally obtained as
\begin{equation}
\hat{\bar{\mathbf{c}}}_1
=
\frac{\hat{\mathbf{c}}_1}{\|\hat{\mathbf{c}}_1\|_2},
\qquad
\hat{\bar{\mathbf{c}}}_2
=
\frac{\hat{\mathbf{c}}_2}{\|\hat{\mathbf{c}}_2\|_2}.
\label{eq:dual_source_pol_estimates}
\end{equation}

To summarize, our dual-source algorithm consists of detecting the strongest source, subtracting its estimated contribution, detecting the weaker source from the residual, and finally refitting both polarization vectors jointly for the two estimated directions.

\section{Algorithm for Optimizing the DMA Configuration Sequence}
\label{sec:OptimAlgorithm}

We now use the calibrated MNT model to optimize the sequence of DMA configurations used for DoA-P estimation. We expect the optimization to be most relevant in the low-$K$, intermediate-SNR regime. When only a small number of configurations is available, the randomly sampled response patterns may not span the direction-polarization dictionary sufficiently well, so the choice of configurations can strongly affect the inverse problem. At very low SNR, however, the estimation accuracy is dominated by noise. Conversely, at very high SNR, even weakly conditioned measurement directions may remain exploitable. We therefore expect the largest benefit of optimized configuration sequences when $K$ is small or moderate and the SNR is neither too low nor too high.

The key enabler for this optimization is the experimentally calibrated MNT model for our specific DMA prototype. Indeed, the calibrated MNT model provides an accurate forward map from \textit{any admissible} binary configuration to the dual-polarized far-field pattern of \textit{our specific DMA prototype}, allowing us to optimize over arbitrary binary configurations without additional radiation-pattern measurements. Without such a forward map, efficiently optimizing the DMA configuration sequence would be very difficult. On the one hand, model-agnostic  measurement-in-the-loop optimization strategies, such as those used in~\cite{del2019optimally,del2020optimal}, are prohibitively time-consuming in our present context because they would require one full far-field radiation pattern measurement per iteration. On the other hand, a feasible model-agnostic approach would consist of selecting a subset from a larger set of random DMA configurations for which the far-field radiation pattern has been measured~\cite{li2024measurement,zhao2026EuCAP}, but this approach drastically limits the design space. Hence, an efficient optimization of the DMA configuration sequence requires an accurate forward map in our present context. Earlier theoretical works relied on a known physics-consistent coupled-dipole model that structurally resembles the MNT model~\cite{del2020learned,qian2022noise}; now, thanks to recent progress in experimentally calibrating the MNT model~\cite{tapie2026experimental,tapie2026channel}, we can apply model-based optimization strategies to concrete experimentally available DMA prototypes.

Ultimately, the goal of our optimization is to minimize the DoA-P error. In our optimization, we use a computationally inexpensive surrogate objective, because the objective must be evaluated repeatedly during optimization. We consider three distinct surrogate objectives. To formalize our surrogate objective formulation, we define the sensing matrix associated with a candidate sequence
$\mathcal{V}=\{\mathbf{v}_1,\mathbf{v}_2,\ldots,\mathbf{v}_K\}$. Using our calibrated MNT model, we can predict the corresponding far-field response vectors
$\mathbf{h}_k=\mathbf{h}(\mathbf{r}(\mathbf{v}_k))$.
To reduce the computational burden during optimization, we evaluate the objective on a subset
$\mathcal{D}\subset\{1,\ldots,N_\mathrm{D}\}$
of $N_{\mathrm{D,opt}}=|\mathcal{D}|$ directions.\footnote{In practice, we retain every tenth direction of the full angular grid and exclude directions within $3^\circ$ of the poles, where the spherical polarization basis becomes ill-conditioned and neighboring azimuthal samples are nearly redundant.} We denote by $\mathbf{h}_{\mathcal{D},k}\in\mathbb{C}^{2N_{\mathrm{D,opt}}}$ the subvector of $\mathbf{h}_k$ containing the $E_\phi$ and $E_\theta$ responses over all directions in $\mathcal{D}$ for the $k$th DMA configuration. The corresponding sensing matrix is
\begin{equation}
\mathbf{H}_{\mathcal{D}}(\mathcal{V})
=
\bigl[
\mathbf{h}_{\mathcal{D},1},
\ldots,
\mathbf{h}_{\mathcal{D},K}
\bigr]^\top
\in\mathbb{C}^{K\times 2N_{\mathrm{D,opt}}}.
\label{eq:optimization_sensing_matrix}
\end{equation}

Our first surrogate objective is the effective rank~\cite{roy2007effective} of the raw sensing matrix (analogous to the optimization objective used in~\cite{del2019optimally,del2020optimal}):
\begin{equation}
R_\mathrm{eff}(\mathbf{H}_{\mathcal{D}})
=
\exp\left(
-\sum_i p_i \log p_i
\right),
\qquad
p_i=\frac{\sigma_i}{\sum_j \sigma_j},
\label{eq:effective_rank}
\end{equation}
where $\sigma_i$ denotes the $i$th singular value of $\mathbf{H}_{\mathcal{D}}$. A flatter singular-value spectrum yields a larger effective rank and indicates that the selected configurations provide more diverse measurements over the considered direction-polarization dictionary. 

Our second surrogate objective is the effective rank after column normalization of the sensing matrix. Specifically, before computing the effective rank, each column of $\mathbf{H}_{\mathcal{D}}$ is normalized to unit $\ell_2$ norm. This removes the relative gains of individual direction-polarization columns and focuses the objective on the diversity of their configuration-dependent response patterns.

Our third surrogate objective is the effective rank after direction-block normalization. Here, for each candidate direction, the two columns corresponding to the $E_\phi$ and $E_\theta$ responses are normalized jointly to unit Frobenius norm. This removes the overall gain associated with that direction while preserving the relative balance between the two transverse polarization components. This normalization is motivated by the structure of the DoA-P estimator in Sec.~\ref{sec:algorithms}: multiplying both columns of $\mathbf{D}_d$ by the same nonzero scalar does not change the projection score $\eta(d)$, because the unknown source strength and propagation loss are absorbed into the fitted polarization-and-amplitude vector. 

For each of the three objectives in turn, we optimize a binary configuration sequence using a multi-start greedy coordinate-ascent algorithm. The choice of this optimization strategy is motivated by the findings in~\cite{hammami2026statistical}. Each of the $K N_\mathrm{M}$ binary control bits is treated as one coordinate. For each random restart, the sequence $\mathcal{V}$ is initialized with independent random binary entries. We then perform up to a fixed number of full sweeps over all coordinates in random order. For each coordinate, we tentatively flip the corresponding bit, use the calibrated MNT model to recompute only the affected row of $\mathbf{H}_{\mathcal{D}}(\mathcal{V})$, and evaluate the chosen surrogate objective. The flip is accepted only if it increases the objective. The sweep procedure stops early if no bit flip is accepted during a full sweep. The best sequence over all random restarts is retained.

\section{Results}
\label{sec:Results}

\subsection{DMA Design}
\label{subsec:DMAdesign}

The design of our DMA prototype closely follows the one proposed in~\cite{sleasman2020implementation} except that it has eight feeds instead of a single feed. The seven additional feeds are necessary for the MNT model parameter estimation, as discussed in~\cite{tapie2026experimental}; during DoA-P estimation, only the central feed is used while the other seven feeds are left open-circuited. Photographic and schematic images of our DMA prototype are shown in Fig.~\ref{Fig1}a. The radiating elements of our DMA are 96 1-bit-programmable cELC resonators, each parametrized by a PIN diode; the meta-element design is based on~\cite{yoo2016efficient}. In our prototype, all 96 meta-elements have the same design and orientation along the $x$-axis shown in Fig.~\ref{Fig1}b. The 96 meta-elements are coupled to the eight feeds via a quasi-2D chaotic cavity. This cavity results in strong all-to-all coupling, which we deliberately seek because strong mutual coupling boosts the wave-domain flexibility~\cite{prod2025mutual,prod2025benefits}. Further technical details about the implementation of our specific DMA prototype can be found in~\cite{tapie2026experimental,tapie2026channel}.

\subsection{Experimental MNT Model Calibration}

To experimentally calibrate the MNT model for our DMA prototype shown in Fig.~\ref{Fig1}, we closely follow~\cite{tapie2026experimental,tapie2026channel}. As proposed in~\cite{tapie2026experimental}, we first estimate a proxy for the mutual-coupling matrix $\mathbf{\Gamma}$ and the reflection coefficients $\alpha$ and $\beta$ of the two available load states, based on measurements of the $8\times 8$ reflection matrix at the DMA's feeds, using the estimation technique from~\cite{del2025experimental}. Then, we measure the dual-polarized far-field radiation pattern for 220 random DMA configurations (upon excitation from the central feed while the other seven feeds are open-circuited). We adjust our proxy for $\mathbf{\Gamma}$ to account for scattering by the seven open-circuited feeds, as explained in~\cite{tapie2026channel}, and we determine matching proxies for the remaining model parameters $\mathbf{h}_0$, $\mathbf{A}$ and $\mathbf{b}$ using the BTALS algorithm presented in~\cite{tapie2026channel}. 
The far-field radiation pattern is sampled on a spherical angular grid with a $3^\circ$ spacing in azimuthal angle $\phi$ and polar angle $\theta$, comprising $N_\mathrm{D}=7260$ directions. 
To quantify the accuracy of our proxy MNT model, we evaluate the normalized mean squared error (NMSE) for the radiation patterns associated with 30 unseen random DMA configurations. We obtain an NMSE of $-30.5\ \mathrm{dB}$, which is comparable to the NMSE achieved in~\cite{tapie2026channel}. Further details on the estimation of our proxy MNT parameters can be found in~\cite{tapie2026experimental,tapie2026channel}.

\subsection{DMA Characterization}

\begin{figure}
    \centering
    \includegraphics[width=0.8\textwidth]{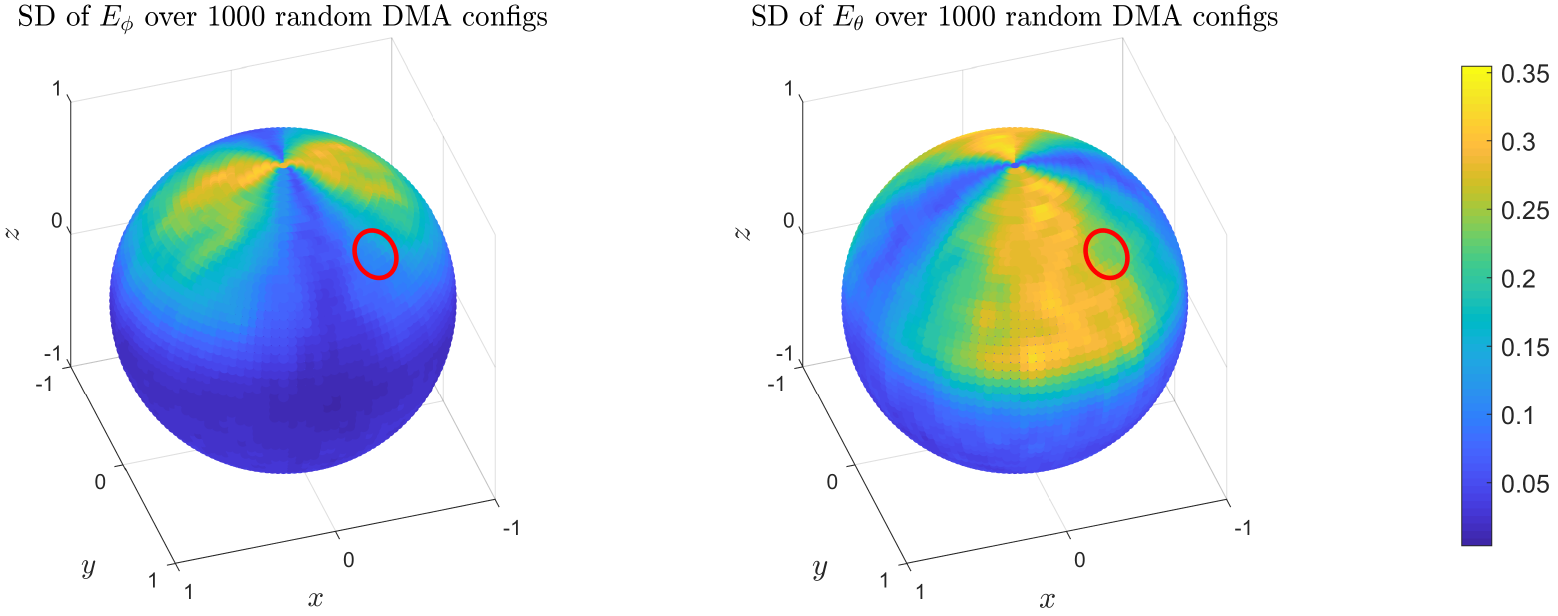}
    \caption{
    Configurational diversity of the DMA's far-field radiation pattern. 
    The maps show the standard deviation (SD) of the two transverse far-field components, $E_\phi$ and $E_\theta$, over 1000 random DMA configurations (evaluated based on the calibrated MNT model). 
    The red circles mark the specific direction used in Fig.~\ref{Fig4}.
    }
    \label{Fig2}
\end{figure}

Before analyzing our ability to optimize the DMA configuration sequence and estimate DoA-P information, we briefly examine our control over the DMA's radiation pattern in this section. To this end, we plot the standard deviation of the two transverse components of the radiated field across 1000 random DMA configurations (evaluated based on our calibrated MNT model). The plotted metric quantifies how strongly the DMA configuration influences the radiation pattern in a given direction and polarization. Clearly, a larger influence improves the ability to leverage the DMA's configurational diversity for computational wireless sensing.
The results displayed in Fig.~\ref{Fig2} reveal a pronounced dipole-like pattern. This behavior is consistent with the common orientation of all programmable meta-elements along the $x$-axis. While this property is not ideal for DoA-P sensing, we defer the design of different DMA architectures with non-uniformly oriented meta-elements to future work.

\subsection{Effective Rank Optimization Outcomes}

\begin{figure}
    \centering
    \includegraphics[width=0.6\textwidth]{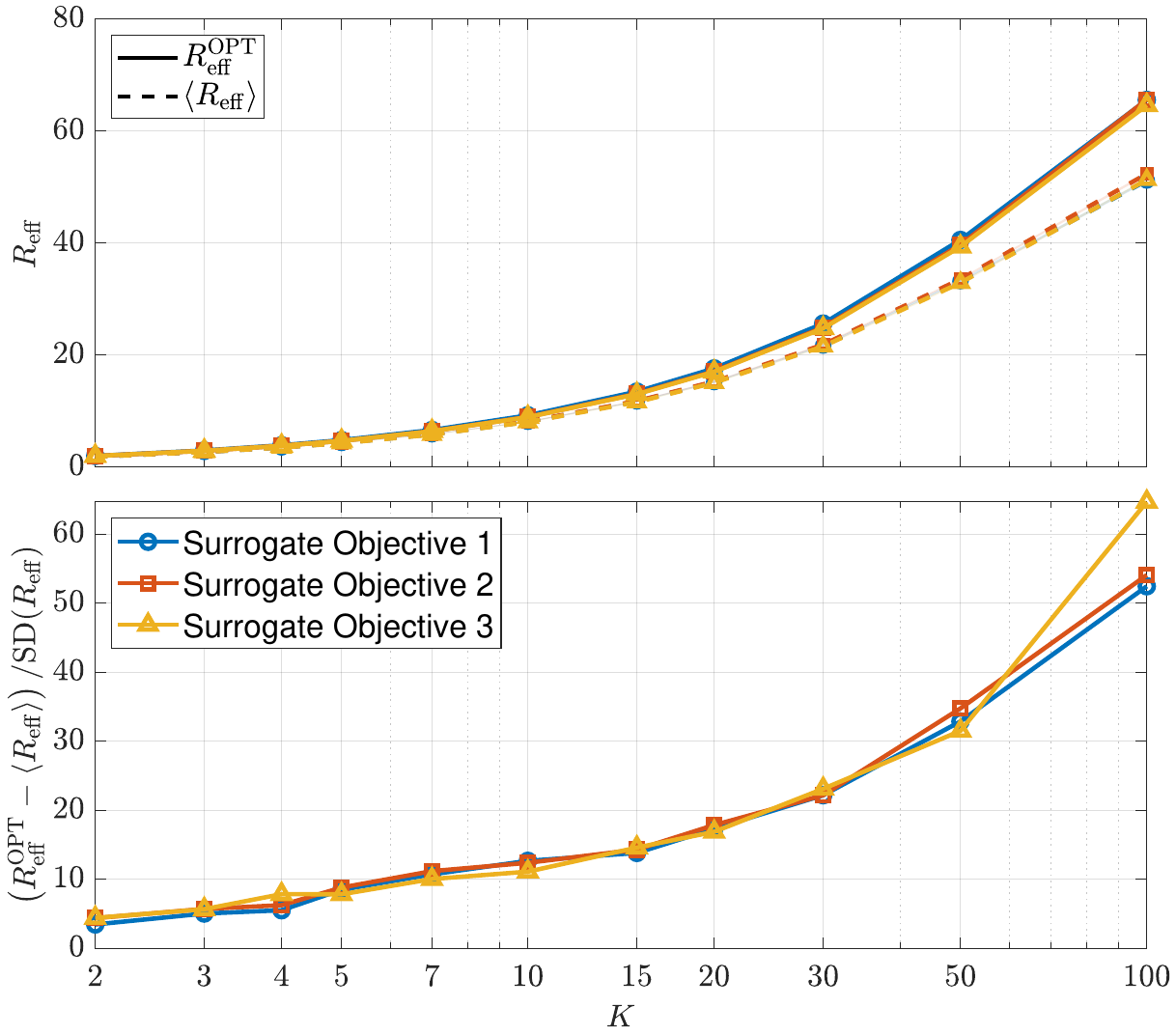}
    \caption{
Effective-rank optimization of the DMA configuration sequence for DoA-P estimation.
Top: $R_\mathrm{eff}$ vs. $K$ for the three surrogate objectives. Solid curves show the optimized values, dashed curves show the mean over 100 random DMA configuration sequences, and shaded regions indicate one standard deviation around the mean. Bottom: optimized-value improvement over the random mean, normalized by the standard deviation over the same 100 random sequences, as a function of $K$. 
}
    \label{Fig3}
\end{figure}

We visualize the outcomes of our surrogate-objective optimizations detailed in Sec.~\ref{sec:OptimAlgorithm} in Fig.~\ref{Fig3}.
For all considered values of $K$, we observe that the optimized sequences outperform the mean random baseline. The standard deviation around the random mean is very small on the scale of the top panel of Fig.~\ref{Fig3}, showing that random configuration sequences yield highly reproducible effective ranks. As a result, even moderate absolute improvements correspond to a clear separation from random configuration sequences. 
We quantify this separation in the bottom panel of Fig.~\ref{Fig3}. The optimized effective ranks lie several standard deviations above the random mean for all $K$, and this normalized improvement increases strongly with $K$. We also observe that the three surrogate objectives yield very similar curves. Overall, with respect to our surrogate objectives, our calibrated MNT-model-based optimizations thus achieve a statistically significant improvement over random configuration sequences.

\subsection{Definitions for DoA-P Performance Evaluation}

\subsubsection{DoA Error}

We quantify the DoA error by the angular separation between the true and estimated unit direction vectors:
\begin{equation}
\epsilon_\mathrm{DoA}
=
\cos^{-1}
\left(
\mathbf{u}_{d_0}^{\top}\mathbf{u}_{\hat d}
\right),
\label{eq:doa_error}
\end{equation}
where $\mathbf{u}_{d_0}$ and $\mathbf{u}_{\hat d}$ are the unit vectors associated with the true and estimated grid directions, respectively.

\subsubsection{Polarization Error}

We quantify the polarization error by the angle between the true and estimated normalized polarization states:
\begin{equation}
\epsilon_\mathrm{pol}
=
\cos^{-1}
\left(
\left|\bar{\mathbf{c}}_0^\dagger\hat{\bar{\mathbf{c}}}\right|
\right),
\label{eq:pol_error}
\end{equation}
where $(\cdot)^\dagger$ denotes the conjugate transpose. The absolute value makes this metric invariant to an arbitrary global complex phase of the estimated polarization vector.

\subsubsection{Signal-to-Noise Ratio (SNR)}

We model the measurement noise in \eqref{eq:single_source_measurement_scalar} as independent circular complex Gaussian noise $n_k \sim \mathcal{CN}(0,\sigma_n^2)$, where $\sigma_n^2=\mathbb{E}\{|n_k|^2\}$ is the noise power per complex scalar measurement. To compare different configuration sequences on an equal footing, we use a fixed absolute noise power for all considered configuration sequences, source directions, and polarization states. We define the SNR as $\mathrm{SNR}_\mathrm{dB}=10\log_{10}(P_\mathrm{ref}/\sigma_n^2)$, where $P_\mathrm{ref}$ is the median, over all valid source directions and several randomly drawn normalized polarization states, of the variance, across an independent reference ensemble of random DMA configurations, of the noiseless received signal for a unit-amplitude source (i.e., $\|\mathbf{c}_0\|_2=1$).

\subsection{Single-Source DoA-P Estimation Performance}

\begin{figure}
    \centering
    \includegraphics[width=\textwidth]{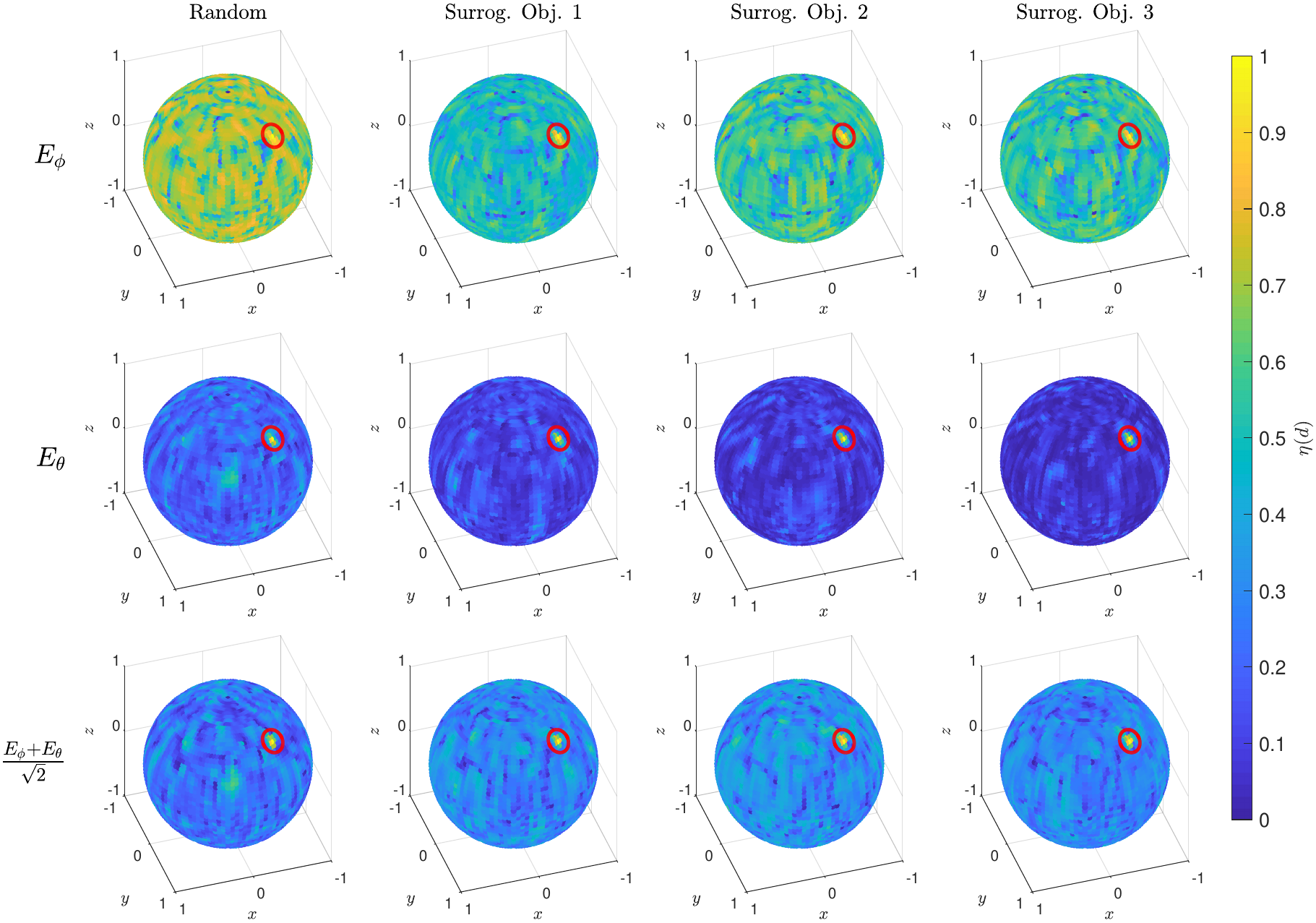}
    \caption{
Maps of $\eta(d)$ in a single-source scenario for three representative source polarizations (rows) and four DMA configuration sequences (columns: random, surrogate objective 1, surrogate objective 2, surrogate objective 3). The maps are shown for $K=20$, $\mathrm{SNR}=25~\mathrm{dB}$, and a source located at $(\phi,\theta)=(120^\circ,45^\circ)$ (marked by the red circle).
}
    \label{Fig4}
\end{figure}

We begin by examining the DoA estimation in a selected single-source scenario. The source direction $(\phi,\theta)=(120^\circ,45^\circ)$ is one of the $N_\mathrm{D}$ directions sampled by our far-field radiation ports. We mark this source direction with a red circle in Fig.~\ref{Fig2}, where we see that in that direction the DMA's configurational diversity is stronger for the $E_\theta$ component than for the $E_\phi$ component. We consider a moderate SNR of 25~dB and a moderate number of measurements of $K=20$. For three representative choices of source polarization, we display the $\eta(d)$ map in Fig.~\ref{Fig4} for four DMA configuration sequences: one random sequence and three optimized sequences, each optimized for one of our three surrogate objectives. Following \eqref{eq:single_source_doa_estimate}, we determine our DoA estimate as the direction in which $\eta(d)$ is largest. Thus, our DoA estimation works best when there is a clear maximum of $\eta(d)$ for the true direction and low background values in other directions.

In all subplots of Fig.~\ref{Fig4}, the global maximum of $\eta(d)$ is aligned with the true source direction, such that we correctly estimate the DoA in all cases. Nonetheless, qualitative differences are evident. The highest background values occur for the $E_\phi$-polarized source, whereas the lowest background values occur for the $E_\theta$-polarized source. This is consistent with Fig.~\ref{Fig2}, since the DMA's configurational diversity is stronger for the $E_\theta$ component in the source direction. Consequently, for a finite noise level, it is easier to distinguish the true direction from incorrect candidate directions for an $E_\theta$-polarized source. In addition, the background level of $\eta(d)$ is higher for the random configuration sequence than for the three optimized sequences. This is consistent with the purpose of the optimization: the optimized sequences reduce spurious projection scores at incorrect directions and thereby improve DoA discrimination. We do not observe clear qualitative differences among the three optimized sequences in Fig.~\ref{Fig4}.

\begin{figure}
    \centering
    \includegraphics[width=\textwidth]{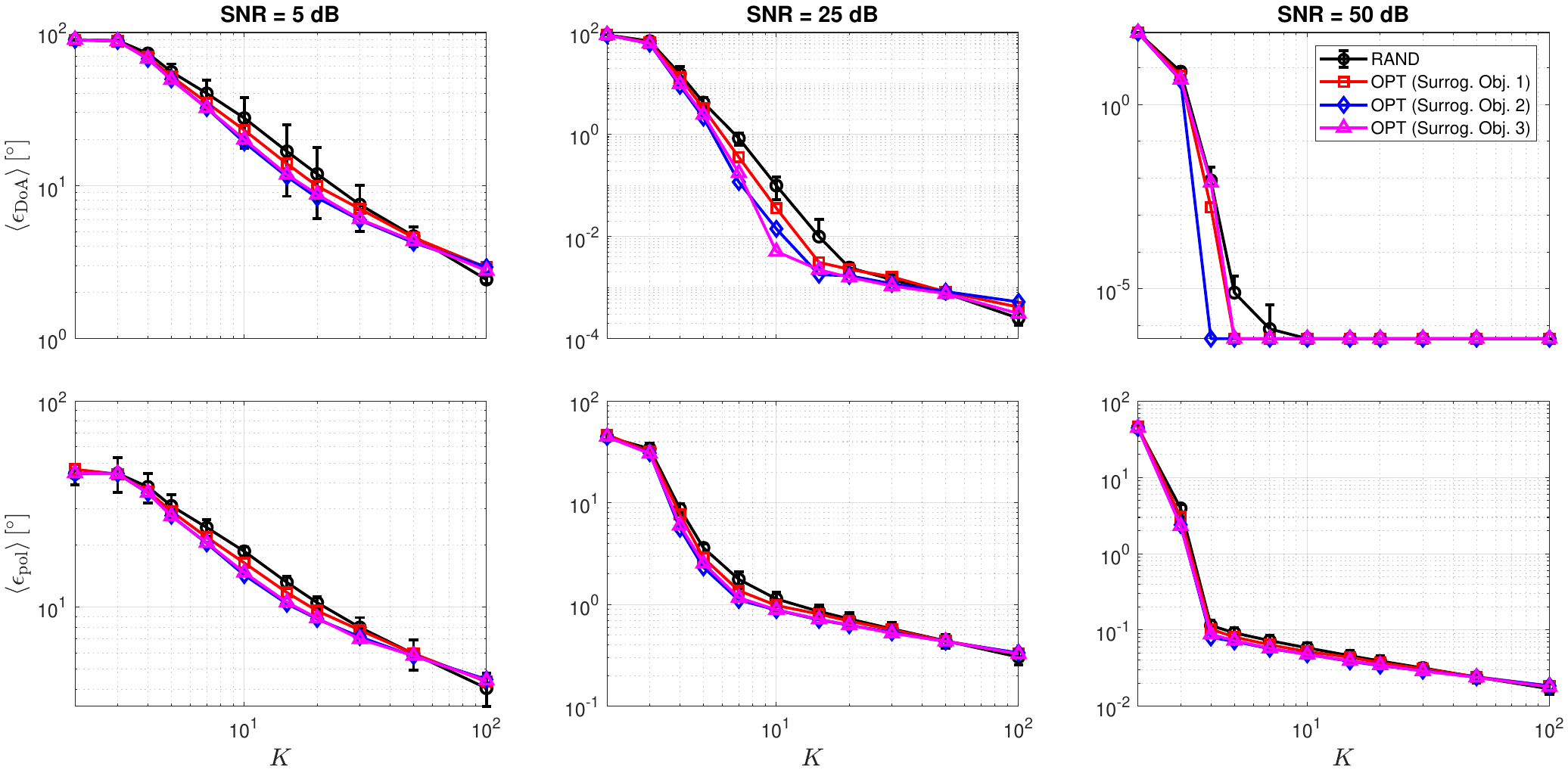}
\caption{
Single-source DoA-P estimation performance for random and optimized DMA configuration sequences.
The mean DoA error $\langle \epsilon_\mathrm{DoA} \rangle$ (top row) and mean polarization error $\langle \epsilon_\mathrm{pol} \rangle$ (bottom row) are shown as a function of $K$ for three SNR values (columns).
}
    \label{Fig5}
\end{figure}

Next, we proceed to a more systematic and quantitative evaluation of the DoA-P estimation performance in the single-source case. For each considered value of $K$, SNR, and DMA configuration sequence, we loop over the three representative normalized polarization states already considered in Fig.~\ref{Fig4} and, for each polarization state, over all non-polar directions of the discretized far-field grid. For each resulting source scenario, we compute the DoA error $\epsilon_\mathrm{DoA}$ and polarization error $\epsilon_\mathrm{pol}$. We then average these errors over all considered source directions and source polarization states. We perform this analysis for the three optimized DMA configuration sequences as well as for 20 random DMA configuration sequences.

The quantitative results in Fig.~\ref{Fig5} confirm that the benefit of our configuration-sequence optimization is most pronounced in the intermediate-SNR, intermediate-$K$ regime. If the DoA and polarization are guessed randomly, the expected errors are $90^\circ$ for $\epsilon_\mathrm{DoA}$ and $45^\circ$ for $\epsilon_\mathrm{pol}$. The former follows because the angular separation between two random directions on a sphere is symmetric about $90^\circ$, while the latter follows because $\epsilon_\mathrm{pol}$ is invariant to a common complex phase, so random normalized complex two-component polarization states are compared over an angular range from $0^\circ$ to $90^\circ$.
Consistently, for very low values of $K$, the averages of $\epsilon_\mathrm{DoA}$  and $\epsilon_\mathrm{pol}$ are very close to these random-guess levels, for all considered SNR levels. 

In the high-SNR case of $\mathrm{SNR}=50~\mathrm{dB}$, as $K$ is increased slightly, the DoA error rapidly drops to a very small value. This small value is possible because the true source positions are restricted to the discretized far-field grid and we only identify the closest grid point rather than estimating a continuous variable. The drop in DoA error is particularly fast and early for the second surrogate objective (column normalization), which reaches the minimum already around $K=4$, whereas the random baseline reaches it only around $K=10$. In contrast, the high-SNR polarization error does not collapse to the same minimum, because the polarization state is not selected from a discrete grid; instead, it exhibits a sharp initial decrease followed by a slower continuous improvement with increasing $K$.

At a medium SNR of $\mathrm{SNR}=25~\mathrm{dB}$, the advantage of optimized configurations is clearest for moderate $K$. For example, around $K=10$, the random baseline yields a mean DoA error of about $0.1^\circ$, whereas the three optimized sequences yield DoA errors of approximately $0.036^\circ$, $0.014^\circ$, and $0.005^\circ$ for surrogate objectives 1, 2, and 3, respectively. For the polarization error at the same SNR, the largest improvement occurs around $K=7$: the random baseline yields a mean polarization error of $1.77^\circ$, while the optimized sequences yield polarization errors of $1.37^\circ$, $1.11^\circ$, and $1.17^\circ$ for surrogate objectives 1, 2, and 3, respectively. For the polarization error, the improvement is more modest than for the DoA error. The ranking of the different surrogate objectives differs between DoA and polarization estimation. 

In the low-SNR case of $\mathrm{SNR}=5~\mathrm{dB}$, optimization still provides some improvement, especially for the second and third surrogate objectives, but the overall errors remain comparatively large even for $K=100$, indicating that the performance is primarily noise-limited. 

Overall, the observed trends in Fig.~\ref{Fig5} agree with the expectation that our surrogate-objective optimization is most useful when the inverse problem is neither dominated by noise nor already essentially solved by high-SNR measurements, and when $K$ lies in an intermediate regime: for very small $K$, there is insufficient information irrespective of the selected configurations, whereas for sufficiently large $K$, even random configuration sequences provide enough diversity to solve the estimation problem reliably.

\subsection{Dual-Source DoA-P Estimation Performance}

Finally, we turn our attention to a dual-source scenario involving a jammer and a desired transmitter. We assume that the jamming signal is much stronger than the desired signal, and our dual-source DoA-P algorithm in Sec.~\ref{subsec:dual_source_algorithm} exploits this received-strength imbalance. We consider a selected example in which the jammer is located at $(\phi,\theta)=(120^\circ,45^\circ)$ and the desired transmitter is located at $(\phi,\theta)=(-40^\circ,45^\circ)$. Both sources transmit with the normalized polarization state $(E_\phi+E_\theta)/\sqrt{2}$ in their respective local spherical bases. The effective strength of the desired transmitter is set 20~dB below the effective strength of the jammer. 

\begin{figure}
    \centering
    \includegraphics[width=0.8\textwidth]{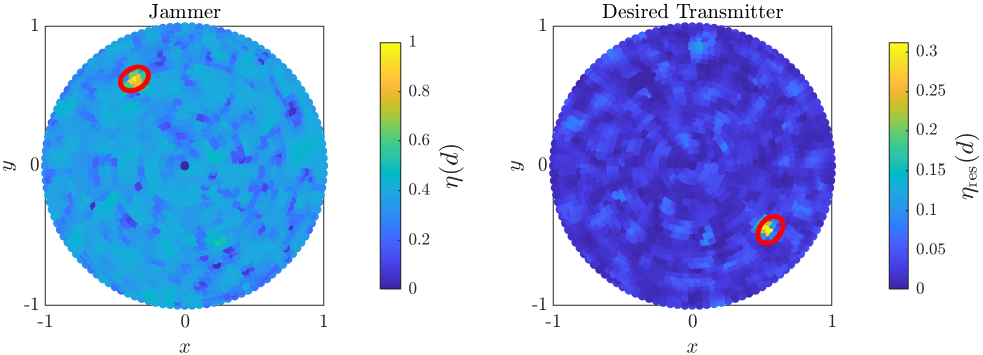}
    \caption{
    Dual-source DoA estimation in the presence of a strong jammer at $(\phi,\theta)=(120^\circ,45^\circ)$ and an effectively 20~dB weaker desired transmitter at $(\phi,\theta)=(-40^\circ,45^\circ)$. 
    The left map of $\eta(d)$ is used to estimate the jammer's DoA while the right map of $\eta_{\mathrm{res}}(d)$ is used to estimate the desired transmitter's DoA. The true DoAs are indicated by a red circle in each case. The results are for $K=100$ configurations optimized with surrogate objective 2 and $\mathrm{SNR}=25~\mathrm{dB}$. 
    }
    \label{Fig6}
\end{figure}

Using a sequence of $K=100$ DMA configurations optimized for the second surrogate objective, and considering a medium SNR level of 25~dB, we correctly identify both DoAs, and we accurately estimate the two polarizations: $\epsilon_\mathrm{DoA}^\mathrm{JAM} = \epsilon_\mathrm{DoA}^\mathrm{DT} = 0.00^\circ$, $\epsilon_\mathrm{pol}^\mathrm{JAM} = 0.25^\circ$, and $\epsilon_\mathrm{pol}^\mathrm{DT} = 1.32^\circ$. While the peak of $\eta_\mathrm{res}(d)$ is only 0.31 in this case, it does allow us to correctly identify the desired transmitter's DoA.

\section{Conclusion}
\label{sec:Conclusion}

To summarize, we have optimized sequences of DMA configurations for DoA-P estimation based on an experimentally calibrated MNT model of a fabricated DMA prototype. The experimentally calibrated MNT model unlocks efficient model-based optimization of DMA configurations for real-world prototypes. We considered three surrogate optimization objectives related to the effective rank of the sensing matrix. The optimized DMA configuration sequences yielded measurable gains in DoA-P estimation performance, especially in the intermediate-SNR and intermediate-$K$ regime. We further extended our DoA-P estimation framework to a dual-source setting involving a weak desired transmitter and a much stronger jammer. While we considered a specific DMA architecture, the MNT formulation is broadly applicable. Therefore, our approach of experimentally calibrating the MNT model and subsequently using it for model-based optimization can be directly applied to other DMA architectures.

Looking forward, our work has revealed avenues for improved DMA architectures, notably designs with diverse meta-element orientations, to provide richer configurational diversity which is expected to improve the DoA-P sensing performance. A further possible extension consists in working with intensity-only data, which would simplify the required receiver hardware. Finally, integrating calibrated-model-based DoA-P sensing with communication protocols could enable DMA-empowered jamming-resilient wireless communications~\cite{yven2025end,hao2025dynamic}.

\begin{backmatter}

\bmsection{Funding}
This work was supported in part by the Nokia Foundation (project 20260028), the ANR France 2030 program (project ANR-22-PEFT-0005), the ANR PRCI program (project ANR-22-CE93-0010), the French Defense Innovation Agency (project 2024600), the European Union's European Regional Development Fund, and the French region of Brittany and Rennes Métropole through the contrats de plan État-Région program (projects ``SOPHIE/STIC \& Ondes'' and ``CyMoCoD'').

\bmsection{Acknowledgment}
The authors acknowledge IETR's QOSC test facility (which is part of the CNRS RF-Net network).

\bmsection{Disclosures}
The authors declare no conflicts of interest.

\bmsection{Data availability}
Data underlying the results presented in this paper are not publicly available at this time but may be obtained from the corresponding author upon reasonable request.

\end{backmatter}

%%%%%%%%%%%%%%%%%%%%%%% References %%%%%%%%%%%%%%%%%%%%%%%%%

\end{document}